# Crystal growth and electronic properties of a 3D Rashba material, BiTeI, with adjusted carrier concentrations

**Manabu Kanou and Takao Sasagawa**

Materials and Structures Laboratory, Tokyo Institute of Technology, Kanagawa 226-8503, JAPAN

E-mail: sasagawa@msl.titech.ac.jp.

**Abstract.** 3D Rashba materials can be a leading player in spin-related novel phenomena, ranging from the metallic extreme (unconventional superconductivity) to the transport intermediate (spin Hall effects) to the novel insulating variant (3D topological insulating states). As the essential backbone for both fundamental and applied research of such a 3D Rashba material, this study established the growth of sizeable single crystals of a candidate compound BiTeI with adjusted carrier concentrations. Three techniques (standard vertical Bridgman, modified horizontal Bridgman, and vapour transport) were employed, and BiTeI crystals ($> 1 \times 1 \times 0.2$ mm$^3$) with fundamentally different electronic states from metallic to insulating were successfully grown by the chosen techniques. The 3D Rashba electronic states, including the Fermi surface topology, for the corresponding carrier concentrations of the obtained BiTeI crystals were revealed by relativistic first-principles calculations.

## 1. Introduction

Manipulating the spin of electrons without external magnetic fields has attracted great attentions in the field of spintronics. Strong spin-orbit interactions (SOI) under the broken space inversion symmetry can induce spontaneous spin polarization without external magnetic fields, which is known as the Rashba effect [1]. This phenomenon has been well studied for the two-dimensional electronic states; e.g. the hetero-interface of semiconductors and the surface of heavy metals [2,3]. In these 2D systems, due to the broken inversion symmetry at the hetero-interface and the surface, electrons feel a potential gradient ($\nabla V$; i.e. electric field $E_z$) perpendicular to the 2D plane. In the presence of the strong SOI, because the potential gradient creates an effective magnetic field, the energy level ($E$) of electrons discernibly depends both on the direction of spins and the electron momentum ($k$), leading to a characteristic spin-splitting of band dispersions without any external magnetic field or spontaneous magnetization. The resulting spin polarized band dispersions are expressed as $E(k)^{\pm} = E_0(k) \pm \alpha \times |k|$, where the first term is the energy state without the Rashba effect and the second one is the spin/momentum dependent energy-shift. The coefficient $\alpha$ is called a Rashba parameter and proportional to the strengths of SOI and $E_z$. The former can be enhanced by incorporating heavier elements into the system, whereas the latter can be relatively easy to design in the case of interfaces and surfaces. Thus, the 2D system has been so far the central focus of the experimental study of the Rashba phenomenon.

 On the other hand, due to the lack of a model compound with a substantially large internal $\nabla V$, the Rashba effect in three-dimensions has been poorly explored. If the gigantic Rashba effect can be induced in the bulk material, it will pave the way for investigating spin-involved novel phenomena relevant to both fundamental physics and technological applications, such as unconventional superconductivity with the spin singlet-triplet hybrid pairing, [4] the intrinsic spin Hall effect, [5,6] and a spin transistors [7].

 In this study, we demonstrate that a ternary compound of bismuth telluroiodide (BiTeI) is a qualified candidate compound as a 3D Rashba material. As shown in figure 1(a), BiTeI is composed of a sequence of Bi, Te, and I layers. This layered crystal structure has the trigonal space group of $P3m1$ [8] without the space inversion symmetry. Thanks to the nearly ionic character of the atomic bonding [8], a considerably large internal $\nabla V$ is expected along the layer stacking direction. In addition, all of the constituent atoms are heavy elements, which should introduce the strong atomic SOI into the compound. Therefore, the required conditions to realize the 3D Rashba effect are fulfilled in BiTeI. Furthermore, the layered crystal structure suggests that BiTeI has an easy cleavage along the trigonal basal planes. Once sizable single crystals (surface area > 1 × 1 mm$^2$) become available, direct observation of the Rashba effect will be possible by scanning tunnelling microscopy/spectroscopy (STM/STS) and angle-resolved photoemission spectroscopy (ARPES) techniques, both of which have been already proved to be powerful in studing of the materials with strong SOI, such as topological insulators [9,10].

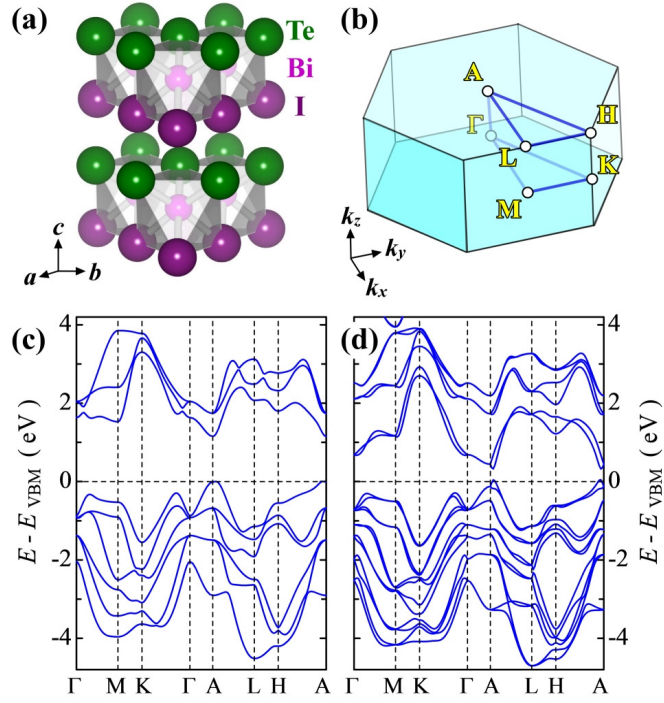

**Figure 1.** (a) The crystal structure of BiTeI, and (b) the corresponding hexagonal Brillouin zone with the symmetry points labelled by the standard notation. Electronic band structure of BiTeI: (c) with spin-orbit interactions (SOI) and (d) without SOI, where vertical energy scale is taken with respect to the valence band maximum (VBM).

Recently, it was confirmed by spin-ARPES measurements that BiTeI had, in fact, spin-split bands with a giant Rashba parameter of $\alpha$ = 3.8 eV Å [11]. Although the observation of the electronic structures in BiTeI was successful in this recent study, there is much room for improvement, in particular with regard to the material aspect of the targeted system, which includes compositions, carrier density, and the quality of single crystals. Without incorporating any foreign element (such as magnetic Mn [11]), high-quality sizeable single crystals of BiTeI were prepared in this study, with a successful adjustment of their electronic states, for the first time, from metallic to insulating ones by the choice of the growth techniques. Together with the experiments, the first-principles calculations taking into account the SOI were performed to infer the detailed electronic structures realized in our crystals.

## 2. Experimental

### 2.1. Crystal growth

The single crystals of BiTeI were grown by standard the vertical Bridgman (VB), the modified horizontal Bridgman (HB), and the vapour transport (VT) techniques. In all methods, stoichiometric amounts of high purity (5Nup) elemental Bi, Te, and I were put in a quartz tube.

It was evacuated and sealed, while being cooled by liquid nitrogen to prevent iodine from sublimation during the process. Before the crystal growth, the sealed quartz tube was heated up to 900 °C in order to completely react the mixture of the starting materials, resulting in a silver ingot of the BiTeI precursor. The VB growth was carried out using a vertical two-zone furnace with settings of 800 °C for the upper zone, 650 °C for the lower zone, and 1 mm h$^{-1}$ downward movement of the sealed quartz tube across the two zones. In the HB technique, the sealed quartz tube was placed in a horizontal two-zone furnace in such a way that the BiTeI precursor ingot was located between the two heating zones. After both zones were heated to 800 °C, one zone was decreased to 750 °C. With this temperature difference of 50 °C maintained, both zones were slowly cooled to room temperature over 50 h. The VT technique utilized a home-made transparent two-zone furnace with the temperature minimum in between the two heating units [12]. The BiTeI precursor ingot in the sealed quartz tube was located at the higher temperature zone. The growth temperature was optimized by the real time observation of the crystal nucleation region using an optical microscope.

*2.2. Characterizations*

The crystal structure of the obtained samples was determined using an X-ray diffractometer (Bruker: D2 PHASER) equipped with a 1D-detector (Bruker: LynxEye). The electrical transport properties of the crystals, including resistivity, the Seebeck coefficient, and the Hall coefficient, were evaluated by the conventional techniques using a home-built setup with a closed-circle He refrigerator. The optical reflectivity spectra between 0.05 and 0.9 eV were measured using a Fourier transform-type interferometer (JASCO: FT/IR-6100) equipped with an infrared microscope (JASCO: IRT-1000).

*2.3. Theoretical calculations*

First-principles calculations in the framework of density functional theory (DFT) were performed within the generalized gradient approximation (GGA), using the full potential linearized augmented plane wave (LAPW) method as implemented in WIEN2k code [13]. Within the experimental unit cell ($a$ = 4.340 Å and $c$ = 6.854 Å), the atomic coordinates were optimized based on the minimum of the force on each atom. Spin-orbit coupling was taken into account as a second variational procedure. The Brillouin zone was sampled with the Monkhorst-Pack scheme, [14] with the momentum grids finer than $\Delta k$ = 0.05 Å$^{-1}$ (e.g. 0.02 Å$^{-1}$ for visualization of the Fermi surface). The electronic band dispersions were calculated along the high symmetry lines of the Brillouin zone, which is shown in figure 1(b) with the standard notation of the symmetry points.

## 3. Results and discussion

### 3.1. Band structure of stoichiometric BiTeI

The calculated electronic band structures for BiTeI before and after taking into account the SOI are shown in figures 1(c) and (d), respectively, and agree with the recent report [15]. The energy level is taken with respect to the valence band maximum (VBM; which is the top of the filled band). The bands within 5 eV below the VBM energy are mainly composed of the 5p-orbitals (6 spin-degenerate bands) of Te and I, while those above 4 eV are of the 6*p*-orbitals (3 bands) of Bi. This is in good agreement with the ionic model of BiTeI, and then the corresponding closed-shell configurations of the all constituent atoms ($Bi^{3+}$, $Te^{2-}$, and $I^-$) account well for the existence of an insulating band gap in BiTeI. As seen in figure 1(d), the number of the bands is markedly increased once the SOI are included in the calculations. Bands are found to be clearly doubled at certain momentum points, indicating that the energy states are no longer degenerated with respect to the spin direction when the SOI are taken into account for BiTeI. This result is a clear sign of the presence of both the intense internal electric field and the strong SOI, and thus of the realization of the 3D Rashba effect in this material.

### 3.2. Crystal growth by Bridgman techniques

Motivated by the results of the theoretical calculations, we have tried to grow sizable single crystals of BiTeI and to control their carrier concentrations in order to explore the novel phenomena arising from the 3D Rashba effect. On the growth of BiTeI single crystals, there have been several reports, [16-18], most of which employ the Bridgman technique using a vertical two-zone furnace. Similar to these reports, we found that the crystals grown by the Bridgman technique always had metallic conductivity with a carrier concentration larger than $n = 5 \times 10^{19}$ cm$^{-3}$ and contained a small amount of the secondary phase of BiI$_3$. By modifying the Bridgman crystal growth from vertical to horizontal, we have succeeded both in reducing the carrier concentration and in suppressing the secondary phase. The Hall resistivity $\rho_{xy}$ at room temperature as a function of the magnetic field in the crystals grown by the vertical and horizontal Bridgman (VB and HB) techniques is shown in the inset of figure 2, of which the slope is inversely proportional to the carrier concentration. The negative slope indicates that both crystals have electron carriers, with a concentration that decreases from $n = 6 \times 10^{19}$ cm$^{-3}$ in the VB crystal to $9 \times 10^{18}$ cm$^{-3}$ in the HB crystal. In accord with the reduced carrier concentration, the in-plane resistivity $\rho_{xx}$ at room temperature becomes ~7 times larger in the HB crystal than in the VB crystal (figure 2).

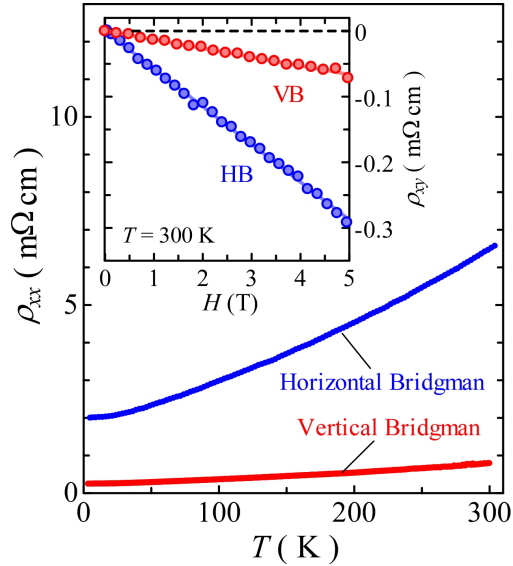

**Figure 2.** Electrical transport property of BiTeI crystals grown by the standard vertical Bridgman (VB) technique and the modified horizontal Bridgman (HB) technique: (main) the temperature dependence of the in-plane resistivity $\rho_{xx}$, (inset) the field dependence of the Hall resistivity $\rho_{xy}$ at room temperature.

The reason for our success might be accounted for as follows. BiTeI has a very large vapour pressure above 300 °C, which is well below its melting point of ~550 °C [19]. Therefore, the solid, liquid, and gas phases coexist during the crystal growth of BiTeI in the Bridgman technique. In case of the VB growth, the solid and liquid interface and the liquid and gas interface are distanced from each other at different temperatures. This will lead to the formation of the secondary phase and enhance the non-stoichiometry in the grown crystals which brings the carriers into the system (i.e. the self-doping effect). On the other hand, in our modified HB technique, the crystal growth proceeds under equilibrium of all three phases at the same temperature, resulting in a suppressed non-stoichiometry and self-doping effect.

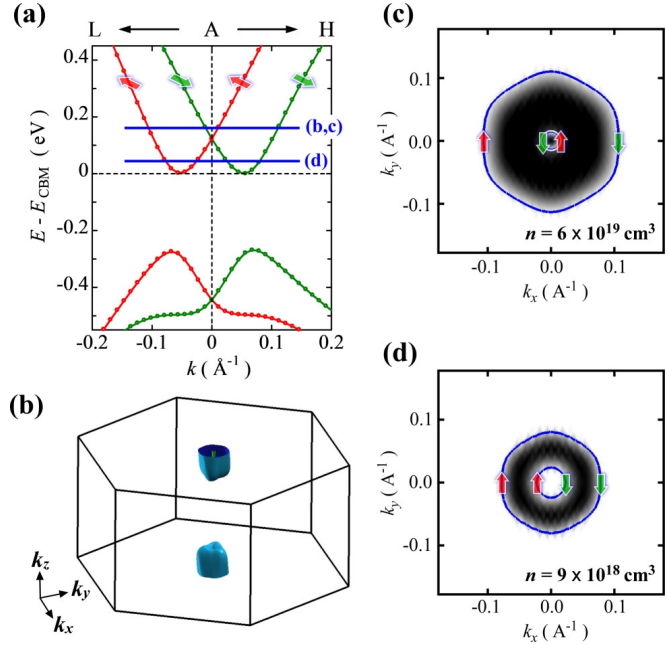

**Figure 3.** (a) Electronic dispersion (energy versus momentum relationship) in BiTeI around the A point, in which the Fermi energy deduced from the measured carrier concentrations in the VB and HB crystals is also indicated by blue lines. The Fermi surfaces of BiTeI in the VB crystal as a 3D rendering (b) and as a 2D cross section at the $k_z = \pi/c$ plane around the A point (c). (d) The 2D Fermi Surface at the $k_z = \pi/c$ plane in the HB crystal.

*3.3. Electronic structure of self-doped BiTeI crystals*

To quantitatively relate the evaluated carrier concentrations in our BiTeI crystals and the detailed electronic structures, the band dispersions near Fermi energy and the Fermi surface for the corresponding carrier concentrations are derived by the first-principles calculations. As seen in figure 1(d), at a low level of the electron doping, electronic states are predominantly formed around the A point [$k = (0, 0, \pi/c)$] in the Brillouin zone. Therefore, the electronic dispersions around the A point (along the L-A-H lines) are plotted in figure 3(a). The energy level is taken with respect to the bottom of the conduction band (CBM: conduction band minimum). Spin directions are within the $k_x$-$k_y$ plane, perpendicular to the momentum direction, and antiparallel to each other for the red and green dispersions. The two blue lines in figure 3(a) specify the Fermi energy for our crystals grown by the VB and HB techniques, respectively. In the VB crystal ($n = 6 \times 10^{19}$ cm$^{-3}$), the Fermi energy is beyond the Rashba band crossing energy of $E - E_{CBM} \sim 0.12$ eV at the A point. The corresponding 3D Femi surface in the whole Brillouin zone and the cross section at the $k_z = \pi/c$ plane around the A point are shown in figures 3(b) and (c), respectively. Figure 3(c) also includes the spin directions as the arrows. It is noted that overall features (i.e. a pair of hexagonal and circular Fermi surfaces with clockwise and anticlockwise spin helicity) in figure 3(c) are in good agreement with the recent report from spin-ARPES experiments [11]. As the Fermi energy goes below the band crossing energy, the Fermi surface changes its topology together with the spin orientation, which is exactly realized in our crystal grown by the HB technique ($n = 9 \times 10^{18}$ cm$^{-3}$ and $E - E_{CBM} \sim 0.04$ eV) as shown in figure 3(d). To the best of our knowledge, this is the first compound having a doughnuts-like 3D Fermi surface with the helical spin texture on it, which will be an ideal platform for the intrinsic spin Hall effect in the 3D electronic state [20].

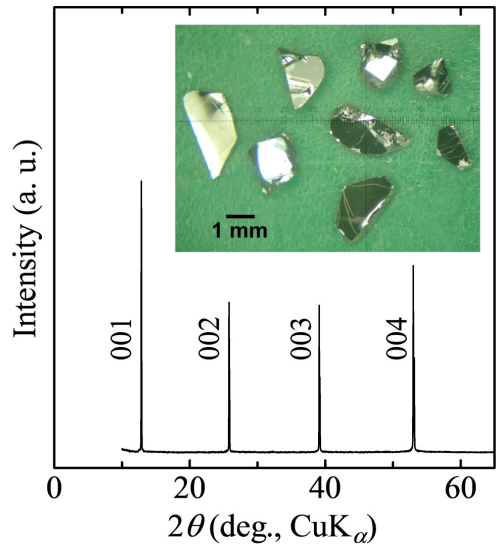

**Figure 4.** X-ray diffraction patterns of BiTeI from a cleaved surface of a single crystal grown by the vapour transport (VT) technique. The inset shows a photograph of the BiTeI single crystals grown by the VT method.

*3.4. Crystal growth by the vapour transport technique*

Considering the high vapour pressure of BiTeI well below its melting point, we thought it possible to adopt a physical vapour transport technique to grow crystals with reduced non-stoichiometry. Like a similar strategy applied to molecular crystals in our previous study [12], the growth process was optimized by the real-time observation using a home-made tube furnace equipped with two transparent heating units. It was found that the vapour pressure was so high above 400 °C that the inside of the quartz tube became opaque due to the thick dark vapour. For a comparable temperature condition to the Bridgman growth (560 °C/470 °C for two zones), the condensate at the nucleation zone was not solid but liquid, which resulted in poly-crystals after cooling. With the aid of real-time observations, it was found out that crystal nucleation took place at temperature settings as low as 300 °C / 250 °C, and sizable single crystals ($> 1 \times 1 \times 0.2$ mm$^3$) were successfully grown in three days (figure 4). The crystals were easily cleaved and flat shiny surfaces were effortlessly obtained. As shown in figure 4, the x-ray diffraction pattern from the cleaved surface has only (00$L$) peaks, indicating that the cleaved surface is parallel to the trigonal basal plane.

To our pleasant surprise, the obtained VT crystals had no electrical conductivity (e.g. resistance larger than 1.5 MΩ) at room temperature. This indicates that the self-doping effect due to non-stoichiometry is negligible. To further confirm their insulating nature, the optical reflectivity was measured on the flat cleaved surface of the VT crystal in addition to the HB crystals using an infrared microscope attached to the Fourier transform-type interferometer (figure 5). In the HB crystal, the reflectivity spectrum has a Drude-like optical plasma edge (extrapolate to 100% of reflectivity at 0 eV) and a reflectivity dip around 0.1 eV, which is consistent with the metallic conductivity having a sufficient carrier concentration. On the other hand, low reflectivity remains down to 0.05 eV in the VT crystal, thus no indication of the free carriers in this compound. Since the ideal stoichiometric BiTeI should become an intrinsic narrow-gap semiconductor, the results suggest that this is realized in our crystals grown by the VT technique. Apart from the BiTeI crystals with the lower carrier density realized by the HB technique, those without carrier by the VT technique provide further potential for spintronics applications, such as the control of both the spins and the carrier concentration by the field effect. Furthermore, at the time of completing this work, a possibility of the topological insulating state without inversion symmetry was theoretically predicted in BiTeI under high pressures [21]. The topological insulator is a novel electronic material, having unusual 2D metallic states on the surface of the insulating bulk. The topological insulator is predicted to stem from strong spin-orbit coupling, and is one of the hottest emerging topics in condensed matter sciences [22]. In this regard, the carrier-free BiTeI crystal made available by the present work also forms the most appropriate starting point for the exploration of such an exotic electronic state.

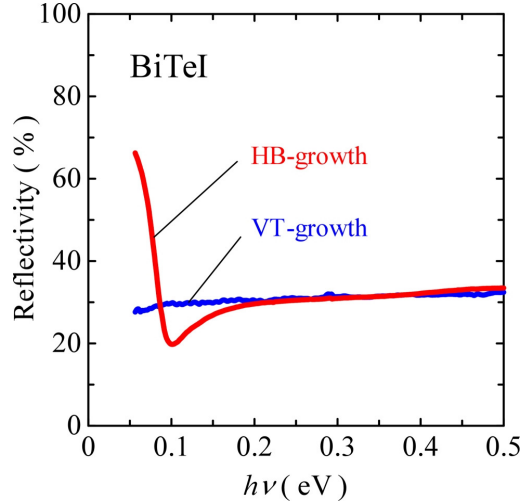

**Figure 5.** Optical reflectivity spectra of BiTeI from cleaved surfaces of single crystals grown by the VT and HB techniques.

## 4. Conclusion

The combination of broken space inversion symmetry, strong atomic spin-orbit interactions, and the nearly ionic character of the atomic bonding makes BiTeI a promising candidate for a 3D Rashba material. Through the realization of sizable single crystals with adjustable from metallic to insulating electronic states, followed by the confirmation of the characteristic spin-split electronic structures by means of relativistic first-principles calculations, this study proved that BiTeI was indeed a qualified 3D Rashba material. Progress on the crystal growth of BiTeI in this study should open up useful pathways for the exploration of novel quantum phenomena, such as unconventional superconductivity, spin transport, and a new type of the topological insulating state, which are anticipated from the unique electronic structure of this compound.

## Acknowledgments

We would like to thank Y. Chen, Y. Kohsaka, S. Murakami, and Z.-X. Shen for fruitful discussions. This work was supported by a Grant-in-Aid for Scientific Research (B) (Grant No.24340078) from MEXT, Japan.